# Solving Power System Differential Algebraic Equations Using Differential Transformation

Yang Liu, *Student Member, IEEE*, Kai Sun, *Senior Member, IEEE*

*Abstract*— This paper proposes a novel non-iterative method to solve power system differential algebraic equations (DAEs) using the differential transformation, which is a mathematical tool able to obtain power series coefficients by transformation rules instead of calculating high order derivatives and has proved to be effective in solving state variables of nonlinear differential equations in our previous study. This paper further solves non-state variables, e.g. current injections and bus voltages, directly with a realistic DAE model of power grids. These non-state variables, nonlinearly coupled in network equations, are conventionally solved by numerical methods with time-consuming iterations, but their differential transformations are proved to satisfy formally linear equations in this paper. Thus, a non-iterative algorithm is designed to analytically solve all variables of a power system DAE model with ZIP loads. From test results on a Polish 2383-bus system, the proposed method demonstrates fast and reliable time performance compared to traditional numerical approaches including the implicit trapezoidal rule method and a partitioned scheme using the explicit modified Euler method and Newton Raphson method.

## I. Introduction

SOVLING power system differential algebraic equations (DAEs) is a fundamental computation task of time domain simulation to assess the transient stability of a power system under contingencies. Traditionally, the differential equations are solved by a numerical integration method, and the algebraic network equations are solved by a numerical iteration method at each integration step [1]-[4]. These methods may suffer from huge computation burdens caused by the iterations at each integration step for the convergence of the network equations and the large number of integration steps to ensure the accuracy and numerical stability of solving the differential equations. Moreover, the computation speed can further deteriorate when system states change significantly, or the system model has strong nonlinearity since the network equation is more difficult or even fails to converge by numerical iteration methods.

Many researches are conducted in the literature to accelerate the solving process of power system DAEs, falling into following three categories: 1) model simplification, 2) parallel computing, and 3) semi-analytical methods. In the first category, both the differential equations and algebraic equations of a DAE model can be simplified. For instance, a widely used coherency-based model reduction technique aggregates a group of coherent generators into an equivalent generator [5]-[6]. Also, to avoid solving nonlinear algebraic network equations separately from solving differential equations, many simulation tools assume all constant impedance loads so as to eliminate the network equations of a DAE model and obtain an ordinary differential equation model [3],[7]. However, methods of this category can bring substantial errors in simulation. The second category of methods employ parallel computers to speed up simulation, which decompose the DAE model or computation tasks onto multiple processors such as the Parareal in time method [8]-[9], multi-decomposition approach [10], the domain decomposition method [11]-[12], the waveform relaxation method [13], the instantaneous relaxation method [14]-[15], the multi-area Thevenin equivalent method [16]-[17], and the practical parallel implementation techniques in [7], [18]-[19]. However, these methods still rely on the traditional numerical algorithms to solve DAEs, thus still requiring small-enough integration steps and numerical iterations. Methods of the third category shift some of computation burdens from the online stage to the offline stage, such as the semi-analytical methods recently proposed by [20]-[24] that offline derive approximate, analytical solutions of differential equations for the purpose of online simulation. However, network equations with a DAE model still need to be solved by iterative numerical methods.

In this paper, a novel non-iterative method is proposed to solve the DAE model of a large-scale power system using a differential transformation (DT) method [25]-[26], which has proved to be an effective mathematical tool to solve state variables of power system differential equations in our previous work [27]. First, this paper derives the DTs of the algebraic network equations with current injections. Then, we prove that current injections and bus voltages which are coupled by the original nonlinear network equations, satisfy a formally linear equation in terms of their power series coefficients after DT. Further, a non-iterative algorithm is designed to analytically solve both state variables and non-state variables by power series of time. Simulation results show the proposed method is fast and reliable compared to traditional methods.

The rest of the paper is organized as follows. Section II conceptually describes the proposed method. Section III

This work was supported in part by the ERC Program of the NSF and DOE under NSF Grant EEC-1041877 and in part by NSF Grant ECCS-1610025.
Y. Liu and K. Sun are with the Department of EECS, University of Tennessee, Knoxville, TN 37996 USA (e-mail: yliu161@vols.utk.edu, kaisun@utk.edu).

derives the DTs of the power system DAE model. Section IV designs a non-iterative algorithm using the derived DTs. Section V tests the proposed method on a Polish 2383-bus system. Finally, conclusion is drawn in Section VI.

II. PROPOSED METHOD FOR SOLVING POWER SYSTEM DAE MODEL USING DIFFERENTIAL TRANSFORMATION

A. Introduction of the Differential Transformation

The Differential Transformation (DT) method is an emerging mathematical tool that is developed in applied mathematics and is introduced to the power system field in [27]. For a linear or nonlinear function $x(t)$, its DT $X(k)$ is defined in (1), meaning the $k^{th}$ order power series coefficient of $x(t)$, where $t$ is an independent variable such as time and $k$ is the power series order. A major advantage of the DT method is that it can directly obtain any order power series coefficients by its various transformation rules, without complicated high-order derivative operations.

$$X(k) = \frac{1}{k!}\left[\frac{d^k x(t)}{dt^k}\right]_{t=0} \quad (1)$$
$$x(t) = \sum_{k=0}^{\infty} X(k) t^k$$

B. Conceptual Description of the Proposed Method

A power system DAE model in the state-space representation is given in (2), where $x$ is the state vector, $v$ is the vector of bus voltages, $f$ represents a vector field determined by differential equations on dynamic devices such as synchronous generators and associated controllers, $i$ is the vector-valued function on current injections from all generators and load buses, , and $Y_{bus}$ is the network admittance matrix.

$$\begin{aligned}\dot{x} &= f(x,v)\\ Y_{bus} v &= i(x,v)\end{aligned} \quad (2)$$

In the proposed method, the solution of both state variables and the non-state variables, bus voltages, are approximated by $K^{th}$ order power series in time in (3). The major task is to solve power series coefficients of orders from 0 to $K$. The two steps to obtain these coefficients are conceptually shown below and then elaborated in Sections III and IV, respectively.

$$\begin{aligned}x &= \sum_0^K X(k) t^k\\ v &= \sum_0^K V(k) t^k\end{aligned} \quad (3)$$

1) Step 1: Deriving DTs of Power System DAE Model

The DTs of the DAE model (2) will be derived in Section III and have the general form in (4). Compared with the original DAE model (2), each variable or function $x$, $v$, $f$, $i$ are transformed to their power series coefficients $X(k)$, $V(k)$, $F(k)$, $I(k)$ (denoted by their corresponding capital letters), coupled by a new set of equations in (4). It can be observed that, the left-hand side (LHS) of (4) only contains the $(k+1)^{th}$ order coefficients of state variables and $k^{th}$ order coefficients of bus voltages, respectively, while the right-hand side (RHS) couples $0^{th}$ to $k^{th}$ order coefficients of both state variables and bus voltages by nonlinear functions $F$ and $I$.

$$\begin{aligned}(k+1)X(k+1) &= F(k) = F\big(X(l),V(l)\big), l = 0\cdots k \quad \text{(a)}\\ Y_{bus} V(k) &= I(k) = I\big(X(l),V(l)\big), l = 0\cdots k \quad \text{(b)}\end{aligned} \quad (4)$$

2) Step 2: Solving Power Series Coefficients of State Variables and Bus Voltages

The main task in this step is to solve power series coefficients $X(k)$ and $V(k)$ ($k \geq 1$) from the $(k-1)^{th}$ order coefficients, as indicated by two circled numbers in Fig. 1. Thus, any order coefficients are solvable from $X(0)$ and $V(0)$.

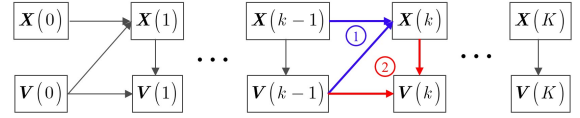

Fig. 1. Recursive process to solve power series coefficients

Rewrite (4a) as (5) by replacing $k$ by $k-1$. Note that $X(k)$ only appears on the LHS and the RHS only contains coefficients up to order $k-1$. Therefore $X(k)$ can be explicitly solved from calculated lower order coefficients.

$$X(k) = \frac{1}{k} F\big(X(l),V(l)\big), l = 0 \cdots k-1 \quad (5)$$

In contrast, from (4b), solving $V(k)$ is not straightforward since it appears on both the LHS and RHS and the vector-valued function $I(\cdot)$ is nonlinear. Later in Section IV, we will prove that the coefficients of current injection $I(k)$ satisfy a formally linear equation (6) about $V(k)$.

$$I(k) = A V(k) + B$$
$$\begin{cases}A = A\big(X(l_1),V(l_2)\big)\\ B = B\big(X(l_1),V(l_2)\big)\end{cases}, \begin{cases}l_1 = 0\cdots k\\ l_2 = 0\cdots k-1\end{cases} \quad (6)$$

Note that the matrices $A$ and $B$ still contain nonlinear functions that only involve the $(k-1)^{th}$ and lower order coefficients on bus voltages, so they do not affect the solvability of $V(k)$. Finally, $V(k)$ is explicitly solved in (7) after substituting (6) into (4b). The detailed derivation of matrices $A$ and $B$ is presented in Section IV.

$$V(k) = (Y_{bus} - A)^{-1} B \quad (7)$$

For complex variables and parameters in (2)-(7) such as current injection vector $i$, bus voltage vector $v$, DTs $I(k)$ and $V(k)$, and admittance matrix $Y_{bus}$, their real and imaginary parts are separate as follows, where $N$ is the number of buses.

$$i = [i_{x,1}, i_{y,1} \cdots i_{x,N}, i_{y,N}]^T$$
$$v = [v_{x,1}, v_{y,1} \cdots v_{x,N}, v_{y,N}]^T$$
$$I(k) = [I_{x,1}(k), I_{y,1}(k) \cdots I_{x,N}(k), I_{y,N}(k)]^T$$
$$V(k) = [V_{x,1}(k), V_{y,1}(k) \cdots V_{x,N}(k), V_{y,N}(k)]^T$$
$$Y_{bus} = \begin{bmatrix} Y_{11} & \cdots & Y_{1N}\\ \vdots & \ddots & \vdots\\ Y_{N1} & \cdots & Y_{NN}\end{bmatrix}, \text{where } Y_{ij} = \begin{pmatrix} G_{ij} & B_{ij}\\ -B_{ij} & G_{ij}\end{pmatrix}$$

**Remark:** There are two important observations: 1) from (6) that current injections and bus voltages, which are coupled by nonlinear network equations in (2), turn out to have linear relationships in terms of their coefficients after DT; 2) coefficients on bus voltages can be explicitly solved by (7) and then used to calculate bus voltages by (3) in a straightforward manner, which is different from using a conventional power flow solver to calculate bus voltages by numerical iterations. The proposed DT based method for solving DAEs differentiates itself from the traditional solution schemes which rely on iterative numerical methods such as the family of Newton Raphson (NR) methods.

## III. DTs OF POWER SYSTEM DAE MODEL

Typically, a power system DAE model contains differential equations for each generator and its controllers, current injection equations for all generator and load buses, and the transmission network equation. DTs of differential equations are provided in [27] and DTs of current injection equations and the network equation are derived in this section.

### A. Vectorized Transformation Rules

In power system DAE model, currents and voltages are usually written as matrix forms using rectangular coordinates. To make the expression of the derived DT more compact, this section extends the existing transformation rules for scalar valued functions to vectorized transformation rules so as to be directly applied to a vector valued function without expanding it into many scalar valued functions first. The proposition 1 provides six vectorized transformation rules in (8) for vector or matrix operations that often appear in a power system DAE model. These rules can be easily obtained by applying the existing transformation rules to each element of the vector valued function and their proofs are omitted.

**Proposition 1:** Given $x(t)$ and $y(t)$ as vector-valued functions having DTs as $X(k)$ and $Y(k)$, $h(t)$ and $H(k)$ as a scalar function and its DT, and $c$ and $d$ are constant matrices, the transformation rules in (8) hold.

$$
\begin{aligned}
&\text{i) } x(t) \pm y(t) \to X(k) \pm Y(k); \quad \text{ii) } x(t)^T \to X(k)^T \\
&\text{iii) } cx(t) \to cX(k); \quad \text{iv) } x(t)d \to X(k)d \\
&\text{v) } x(t)y(t) \to X(k) \otimes Y(k) = \sum_{m=0}^{k} X(m)Y(k-m) \\
&\text{vi) } \frac{1}{h(t)}y(t) \to \frac{1}{H(0)}\left[Y(k) - \sum_{m=0}^{k-1} H(k-m)Z(m)\right]
\end{aligned} \quad (8)
$$

### B. DTs of the Current Injection Equation of Generators

Consider the detailed 6$^{th}$ order synchronous generator model in [27]. The current injection using the $d$-$q$ coordinate system is given in (9). The coordination transformation between $d$-$q$ and $x$-$y$ coordinate system is given in (10). Variables $i_d$, $i_q$ are the $d$-axis and $q$-axis stator currents; $e''_d$, $e''_q$ are $d$-axis and $q$-axis sub-transient voltages; $v_d$ and $v_q$ are the $d$-axis and $q$-axis terminal voltages; $\delta$ is the rotor angle. Parameters $x''_d$, $x''_q$ and $r_a$ are the $d$-axis and $q$-axis sub-transient reactance and internal resistance, respectively.

$$
\begin{bmatrix} i_d \\ i_q \end{bmatrix} = y_a \left( \begin{bmatrix} e''_d \\ e''_q \end{bmatrix} - \begin{bmatrix} v_d \\ v_q \end{bmatrix} \right), \text{ where } y_a = \begin{bmatrix} r_a & -x''_q \\ x''_d & r_a \end{bmatrix}^{-1} \quad (9)
$$

$$
\begin{bmatrix} i_x \\ i_y \end{bmatrix} = r \begin{bmatrix} i_d \\ i_q \end{bmatrix}, \begin{bmatrix} v_x \\ v_y \end{bmatrix} = r \begin{bmatrix} v_d \\ v_q \end{bmatrix}, \text{ where } r = \begin{bmatrix} \sin\delta & \cos\delta \\ -\cos\delta & \sin\delta \end{bmatrix} \quad (10)
$$

The current injection under the $x$-$y$ axis is given in (11) by combining (9)-(10).

$$
\begin{bmatrix} i_x \\ i_y \end{bmatrix} = \tau \begin{bmatrix} e''_d \\ e''_q \end{bmatrix} - \lambda \begin{bmatrix} v_x \\ v_y \end{bmatrix}, \text{ where } \begin{cases} \tau = ry_a \\ \lambda = \tau r^T \end{cases} \quad (11)
$$

The DT of (11) is given in (12).

$$
\begin{bmatrix} I_x(k) \\ I_y(k) \end{bmatrix} = \Gamma(k) \otimes \begin{bmatrix} E''_d(k) \\ E''_q(k) \end{bmatrix} - \Lambda(k) \otimes \begin{bmatrix} V_x(k) \\ V_y(k) \end{bmatrix} \quad (12)
$$

For details of the derivation, the RHS of (12) is obtained using rules i) and v), where the $\Gamma(k)$ and $\Lambda(k)$ are respectively DTs of $\tau$ and $\lambda$, given by rules ii), iv) and v) as follows. $R(k)$ is the DT of the matrix $r$, where $\Phi(k)$ and $\Psi(k)$ are DTs of sine and cosine functions, respectively, given in Proposition 2 in [27].

$$
\tau = ry_a \to \Gamma(k) = R(k)y_a
$$

$$
\lambda = \tau r^T \to \Lambda(k) = \Gamma(k) \otimes R(k)^T
$$

$$
R(k) = \begin{bmatrix} \Phi(k) & \Psi(k) \\ -\Psi(k) & \Phi(k) \end{bmatrix}
$$

Eq. (12) contains the convolution of a 2×2 matrix and a 2×1 vector, and its calculation is the same as the convolution of two matrices in the rule v). The detailed expression is following.

$$
\begin{bmatrix} I_x(k) \\ I_y(k) \end{bmatrix} = \sum_{m=0}^{k} \Gamma(m) \begin{bmatrix} E''_d(k-m) \\ E''_q(k-m) \end{bmatrix} - \sum_{m=0}^{k} \Lambda(m) \begin{bmatrix} V_x(k-m) \\ V_y(k-m) \end{bmatrix}
$$

### C. DTs of the Current Injection Equation of Loads

Consider the ZIP load model [3] in (13) where $p$ and $q$ are the active and reactive power loads, respectively; $v_t$ is the bus voltage magnitude defined in (14) and $u$ equals its square; $p_0$, $q_0$ and $v_{t0}$ are the steady state active power, reactive power and bus voltage; $a_p$ and $a_q$ are the percentages of constant impedance load; $b_p$ and $b_q$ are the percentages of constant current load; and $c_p$ and $c_q$ are the percentages of constant power load. There are $a_p + b_p + c_p = 1$ and $a_q + b_q + c_q = 1$.

$$
\begin{cases}
p = p_0 \left( a_p \left(\dfrac{v_t}{v_{t0}}\right)^2 + b_p \left(\dfrac{v_t}{v_{t0}}\right) + c_p \right) \\
q = q_0 \left( a_q \left(\dfrac{v_t}{v_{t0}}\right)^2 + b_q \left(\dfrac{v_t}{v_{t0}}\right) + c_q \right)
\end{cases} \quad (13)
$$

$$
v_t = \sqrt{u}, \quad u = v_x^2 + v_y^2 \quad (14)
$$

The current injected to the network can be calculated from the active and reactive power injections, and is written in matrix forms in (15) where $\beta_a$, $\beta_b$ and $\beta_c$ are constant matrices.

$$
\begin{bmatrix} i_x \\ i_y \end{bmatrix} = \begin{bmatrix} i_x \\ i_y \end{bmatrix}_z + \begin{bmatrix} i_x \\ i_y \end{bmatrix}_i + \begin{bmatrix} i_x \\ i_y \end{bmatrix}_p
$$

$$
\triangleq \frac{1}{v_{t0}^2} \beta_a \begin{bmatrix} v_x \\ v_y \end{bmatrix} + \frac{1}{v_{t0}} \frac{1}{v_t} \beta_b \begin{bmatrix} v_x \\ v_y \end{bmatrix} + \frac{1}{u} \beta_c \begin{bmatrix} v_x \\ v_y \end{bmatrix} \quad (15)
$$

$$
\beta_a = \begin{bmatrix} p_0 a_p & q_0 a_q \\ -q_0 a_q & p_0 a_p \end{bmatrix}, \beta_b = \begin{bmatrix} p_0 b_p & q_0 b_q \\ -q_0 b_q & p_0 b_p \end{bmatrix}, \beta_c = \begin{bmatrix} p_0 c_p & q_0 c_q \\ -q_0 c_q & p_0 c_p \end{bmatrix}
$$

DTs of $u$ and $v_t$ are derived in [27] and are given in (16)-(17). Then, DTs of the RHS terms in (15) can be obtained using rules in (8) as explained in the following.

$$U(k) = V_x(k) \otimes V_x(k) + V_y(k) \otimes V_y(k) \quad (16)$$

$$V_t(k) = \frac{1}{2V_t(0)} U(k) - \frac{1}{2V_t(0)} \sum_{m=1}^{k-1} V_t(m) V_t(k-m) \quad (17)$$

The first term in RHS of (15) is the current injection of constant impedance load. It is the product of a constant number $1/v_{t0}^2$, a constant matrix $\boldsymbol{\beta}_a$ and a vector valued function $[v_x, v_y]^T$. Therefore, its DT is given in (18) using the rule iii).

$$\begin{bmatrix} i_x \\ i_y \end{bmatrix}_z = \frac{1}{v_{t0}^2} \boldsymbol{\beta}_a \begin{bmatrix} v_x \\ v_y \end{bmatrix} \rightarrow \begin{bmatrix} I_x(k) \\ I_y(k) \end{bmatrix}_z = \frac{1}{v_{t0}^2} \boldsymbol{\beta}_a \begin{bmatrix} V_x(k) \\ V_y(k) \end{bmatrix} \quad (18)$$

The second term in RHS of (15) is the current injection of constant current load with DT in (19). It is transformed by three steps. First, the product of the constant matrix $\boldsymbol{\beta}_b$ and the vector valued function $[v_x, v_y]^T$ is transformed using the rule iii). Then, the division of the vector valued function $\boldsymbol{\beta}_b[v_x, v_y]^T$ and the scalar valued function $v_t$ is transformed using the rule vi). Finally, the product of the constant number $1/v_{t0}$ and the vector valued function $1/v_t \boldsymbol{\beta}_b[v_x, v_y]^T$ is transformed using the rule iii).

$$\begin{bmatrix} i_x \\ i_y \end{bmatrix}_i = \frac{1}{v_{t0}} \frac{1}{v_t} \boldsymbol{\beta}_b \begin{bmatrix} v_x \\ v_y \end{bmatrix} \rightarrow$$
$$\begin{bmatrix} I_x(k) \\ I_y(k) \end{bmatrix}_i = \frac{1}{v_{t0}} \frac{1}{V_t(0)} \left( \boldsymbol{\beta}_b \begin{bmatrix} V_x(k) \\ V_y(k) \end{bmatrix} - \sum_{m=0}^{k-1} V_t(k-m) \begin{bmatrix} i_x(m) \\ i_y(m) \end{bmatrix}_i \right) \quad (19)$$

The third term in RHS of (15) is the current injection of constant power load. It contains the product of a constant matrix $\boldsymbol{\beta}_c$ and a vector valued function $[v_x, v_y]^T$, then divided by a scalar valued function $u$. Similar with the constant current load, its DT is given in (20) using rules iii) and vi).

$$\begin{bmatrix} i_x \\ i_y \end{bmatrix}_p = \frac{1}{u} \boldsymbol{\beta}_c \begin{bmatrix} v_x \\ v_y \end{bmatrix} \rightarrow$$
$$\begin{bmatrix} I_x(k) \\ I_y(k) \end{bmatrix}_p = \frac{1}{U(0)} \left( \boldsymbol{\beta}_c \begin{bmatrix} V_x(k) \\ V_y(k) \end{bmatrix} - \sum_{m=0}^{k-1} U(k-m) \begin{bmatrix} I_x(m) \\ I_y(m) \end{bmatrix}_p \right) \quad (20)$$

Finally, the DT of current injection equation (15) is given in (21) by summing DTs of each term (18)-(20) using the rule i).

$$\begin{bmatrix} I_x(k) \\ I_y(k) \end{bmatrix} = \begin{bmatrix} I_x(k) \\ I_y(k) \end{bmatrix}_z + \begin{bmatrix} I_x(k) \\ I_y(k) \end{bmatrix}_i + \begin{bmatrix} I_x(k) \\ I_y(k) \end{bmatrix}_p \quad (21)$$

### D. DTs of the Network Equation

The network equation is in (22), which couples the current injections of all generators and loads. Its DT is given in (23).

$$i = Y_{bus} v_{bus}$$

$$\begin{bmatrix} i_{x,m} \\ i_{y,m} \end{bmatrix} = \begin{cases} \text{RHS of } (11) \text{ for generator buses} \\ -\text{RHS of } (15) \text{ for load buses} \\ \text{RHS of } (11) - \text{RHS of } (15), \text{ for buses with} \\ \qquad \text{both generators and loads} \end{cases} \quad (22)$$

$$I(k) = Y_{bus} V(k)$$

$$\begin{bmatrix} I_{x,m}(k) \\ I_{y,m}(k) \end{bmatrix} = \begin{cases} \text{RHS of } (12) \text{ for generator buses} \\ -\text{RHS of } (21) \text{ for load buses} \\ \text{RHS of } (12) - \text{RHS of } (21) \text{ for buses with} \\ \qquad \text{both generators and loads} \end{cases} \quad (23)$$

## IV. SOLVING POWER SERIES COEFFICIENTS OF STATE VARIABLES AND BUS VOLTAGES

### A. Linear Relationship Between Current Injection and Bus Voltage in terms of Power Series Coefficients

**Proposition 2**: The transformed current injections in (12) and (21) for generators and loads respectively satisfy equations (24) and (25), which are formally linear.

$$\begin{bmatrix} I_x(k) \\ I_y(k) \end{bmatrix} = A_g \begin{bmatrix} V_x(k) \\ V_y(k) \end{bmatrix} + B_g \quad (24)$$

$$\begin{bmatrix} I_x(k) \\ I_y(k) \end{bmatrix} = A_l \begin{bmatrix} V_x(k) \\ V_y(k) \end{bmatrix} + B_l \quad (25)$$

The proofs are given in Appendix and the detailed expressions of matrices $A_g$, $B_g$, $A_l$ and $B_l$ are in (36) and (40). Using this proposition, current injections of all buses can be written as (6) with $A$ and $B$ in (26).

$$A = diag(A_1, A_2 \cdots A_N), A_n = \begin{cases} A_g, \text{for generator buses} \\ -A_l, \text{for load buses} \\ A_g - A_l, \text{for buses with both} \\ \qquad \text{generators and loads} \end{cases}$$

$$B = \begin{bmatrix} B_1^T, B_2^T \cdots B_N^T \end{bmatrix}^T, B_n = \begin{cases} B_g, \text{for generator buses} \\ -B_l, \text{for load buses} \\ B_g - B_l, \text{for buses with both} \\ \qquad \text{generators and loads} \end{cases} \quad (26)$$

### B. Non-iterative Algorithm to Solve Power Series Coefficients

Following the basic idea in Fig. 1, **Algorithm 1** is further designed to solve power series coefficients of both state variables and bus voltages up to any desired order. Note that all the coefficients are explicitly calculated with no iteration.

---

**Algorithm 1: Solve Coefficients**

**Input**: initial values of state variables and bus voltages $x_0, v_0$

**Output**: any order coefficients $X(k), V(k), k = 0 \cdots K$

**Steps**:

Initialization: $X(0) = x_0, V(0) = v_0$

1. Calculate the matrix $A$
   1.1 Calculate the matrix $A_g$ for generators by (36)
   1.2 Calculate the matrix $A_l$ for loads by (40)

2. Calculate the matrix $(Y_{bus} - A)$ and solve $(Y_{bus} - A)^{-1}$

**for** $k = 1 : K$

3. Solve $X(k)$: $X(k) = \frac{1}{k} F(X(l), V(l)), l = 0 \cdots k-1$ using [27]
   3.1 Solve state variables of governors and turbines by (8) and (10) in [27]
   3.2 Solve state variables of $6^{th}$ order generator model by (17) in [27]
   3.3 Solve state variables of IEEE Type I exciter model by (24) in [27]

4. Calculate the matrix $B$

    4.1 Calculate the matrix $B_g$ for generators by (36)

    4.2 Calculate the matrix $B_l$ for loads by (40)

  5. Solve $V(k)$: $V(k) = (Y_{bus} - A)^{-1} B$

**end**

### C. Extension

This section further discusses the linear relationship among non-state variables for a frequency dependent load model. When considering the impact of frequency changes, the ZIP load model is changed to a set of DAEs in (27), where $\theta$ is the bus voltage angle, $\Delta f$ is the frequency change, $i_l$ is the current injection of the ZIP load model, $i_{l,f}$ is the current injection after considering the frequency change, and $d$ is a constant.

$$\begin{aligned} \dot{\theta} &= \Delta f \\ i_{l,f} &= i_l(1 + d\Delta f) \end{aligned} \quad (27)$$

The DT of (27) are given in (28)-(29). For the DT of the algebraic current injection equation (29), it is also proved to satisfy a formally linear equation in (30) with detailed proofs given in the Appendix.

$$\Theta(k) = \frac{1}{k} \Delta F(k-1) \quad (28)$$

$$I_{l,f}(k) = I_l(k) + dI_l(k) \otimes \Delta F(k) \quad (29)$$

$$I_{l,f}(k) = \begin{bmatrix} A'_l & A'_f \end{bmatrix} \begin{bmatrix} V_l(k) \\ \Delta F(k) \end{bmatrix} + B'_l \quad (30)$$

Two additional variables are introduced, i.e., the state variable $\theta$ and the non-state variable $\Delta f$, and their solutions are also approximated by power series of time in (31), where the coefficients $\Theta(k)$ are solved together with the coefficients of state variables $X(k)$ and coefficients $\Delta F(k)$ are solved together with the coefficients of bus voltages $V(k)$.

$$\theta = \sum_0^K \Theta(k) t^k; \quad \Delta f = \sum_0^K \Delta F(k) t^k \quad (31)$$

## V. CASE STUDY

The proposed method is first illustrated on a 3-machine 9-bus power system. Then, to validate the accuracy, time performance and robustness of the proposed method on solving practical high-dimensional nonlinear DAEs, the 327-machine 2383-bus Polish system [29] with detailed models on generators, exciters, governors, turbines, and ZIP loads are used. In the ZIP load model, the percentages of each component are 20%, 30% and 50% respectively.

Two widely used solution approaches are implemented for comparison [3]: 1) **TRAP-NR** method where the differential equations are algebraized by implicit trapezoidal method (TRAP) first and then solved *simultaneously* with the network equations by Newton Raphson (NR) method. 2) **ME-NR** method using a partitioned scheme where the differential and network equations are *alternatively* solved by explicit modified Euler method (ME) and NR method respectively. The time step length of both the TRAP-NR method and ME-NR method is $1\times10^{-3}$ s, while the proposed method prolongs the time step length to 10 times and still achieves better accuracy.

For a fair comparison, the benchmark result is given by the TRAP-NR method using a small enough time step length of $1\times10^{-4}$ s and errors of the proposed method, the TRAP-NR method and the ME-NR method are calculated by their differences from the benchmark result. Simulations are conducted in MATLAB R2018b on a personal computer with i7-6700U CPU.

### A. Illustration on a 3-machine 9-bus Power System

The 3-machine 9-bus power system in [28] is used to illustrate the proposed method. The system contains generators at buses 1 to 3 equipped with governors and exciters whose differential equations can be found in [27], ZIP loads at buses 5, 6 and 8 and transition buses 4, 7 and 9. Equation (32) shows the network equations on current injections:

$$\begin{bmatrix} i_{x,n} \\ i_{y,n} \end{bmatrix} = \tau_n \begin{bmatrix} e''_{d,n} \\ e''_{q,n} \end{bmatrix} - \lambda_n \begin{bmatrix} v_{x,n} \\ v_{y,n} \end{bmatrix}, \quad \text{if } n = 1,2,3$$

$$\begin{bmatrix} i_{x,n} \\ i_{y,n} \end{bmatrix} = \frac{1}{v_{t0,n}^2} \beta_{a,n} \begin{bmatrix} v_{x,n} \\ v_{y,n} \end{bmatrix} + \frac{1}{v_{t0,n}} \frac{1}{\sqrt{v_{x,n}^2 + v_{y,n}^2}} \beta_{b,n} \begin{bmatrix} v_{x,n} \\ v_{y,n} \end{bmatrix} \quad (32)$$

$$+ \frac{1}{v_{x,n}^2 + v_{y,n}^2} \beta_{c,n} \begin{bmatrix} v_{x,n} \\ v_{y,n} \end{bmatrix}, \quad \text{if } n = 5,6,8$$

$$\begin{bmatrix} i_{x,1} \\ \vdots \\ i_{y,9} \end{bmatrix} = \begin{bmatrix} Y_{11} & \cdots & Y_{19} \\ \vdots & \ddots & \vdots \\ Y_{91} & \cdots & Y_{99} \end{bmatrix} \begin{bmatrix} v_{x,1} \\ \vdots \\ v_{y,9} \end{bmatrix}$$

Apply DT to (27) according to (12), (18)-(21). Coefficients of state variables and bus voltages, i.e. $X(1)$, $V(1)$,..., $X(K)$, $V(K)$, are recursively calculated starting from $X(0)$ and $V(0)$. For illustration purpose, calculation of $X(1)$ and $V(1)$ is explained as follows. First, calculate $X(1)$ by (5) with $k=1$:

$$X(1) = F(X(0), V(0))$$

where detailed equations of $F$ can be found in [27]. Then, write $I(1)$ of all generators and loads as a linear equation about $V(1)$ so as to explicitly solve $V(1)$. For instance, current injections from the constant power load component at bus 5 (i.e. the last term of the second equation in (32)) can be calculated by the following detailed steps. The remaining current injections can be handled in the similar way.

$$\begin{aligned} \begin{bmatrix} I_{x,5}(1) \\ I_{y,5}(1) \end{bmatrix}_p &= \frac{1}{U_5(0)} \left( \beta_{c,5} \begin{bmatrix} V_{x,5}(1) \\ V_{y,5}(1) \end{bmatrix} - U_5(1) \begin{bmatrix} I_{x,5}(0) \\ I_{y,5}(0) \end{bmatrix}_p \right) & (a) \\ &= \frac{1}{U_5(0)} \left( \beta_{c,5} \begin{bmatrix} V_x(1) \\ V_y(1) \end{bmatrix} - 2(V_{x,5}(0)V_{x,5}(1) + V_{y,5}(0)V_{y,5}(1)) \begin{bmatrix} I_{x,5}(0) \\ I_{y,5}(0) \end{bmatrix}_p \right) & (b) \\ &= \frac{1}{U_5(0)} \left( \beta_{c,5} \begin{bmatrix} V_x(1) \\ V_y(1) \end{bmatrix} - \begin{bmatrix} 2V_{x,5}(0)I_{x,5}(0) & 2V_{y,5}(0)I_{x,5}(0) \\ 2V_{x,5}(0)I_{y,5}(0) & 2V_{y,5}(0)I_{y,5}(0) \end{bmatrix} \begin{bmatrix} V_{x,5}(1) \\ V_{y,5}(1) \end{bmatrix} \right) & (c) \\ &= \frac{1}{U_5(0)} \left( \beta_{c,5} - \begin{bmatrix} 2V_{x,5}(0)I_{x,5}(0) & 2V_{y,5}(0)I_{x,5}(0) \\ 2V_{x,5}(0)I_{y,5}(0) & 2V_{y,5}(0)I_{y,5}(0) \end{bmatrix} \right) \begin{bmatrix} V_{x,5}(1) \\ V_{y,5}(1) \end{bmatrix} + \begin{bmatrix} 0 \\ 0 \end{bmatrix} & (d) \\ &= \begin{bmatrix} -0.8771 & 0.1566 \\ 0.1566 & 0.8771 \end{bmatrix} \begin{bmatrix} V_{x,5}(1) \\ V_{y,5}(1) \end{bmatrix} + \begin{bmatrix} 0 \\ 0 \end{bmatrix} \triangleq A_{p,5} \begin{bmatrix} V_{x,5}(1) \\ V_{y,5}(1) \end{bmatrix} + B_{p,5} & (e) \end{aligned} \quad (33)$$

For load bus 5, the first order coefficient $[I_{x,5}(1), I_{y,5}(1)]_p^T$ for constant power load is given in (33a) from (20). After substituting the expression of $U(1)$ given in (16) and simple matrix operations, it turns to (33b) and (33c), respectively. Then, all terms containing $[V_{x,5}(1), V_{y,5}(1)]^T$ are combined as a group versus the remaining as another group in (33d). Since $\beta_{c,5}$ is a constant matrix and all variables, $U_5(0)$, $V_{x,5}(0)$, $V_{y,5}(0)$, $I_{x,5}(0)$, $I_{y,5}(0)$, except for $[V_{x,5}(1), V_{y,5}(1)]^T$ have been known, their values are directly substituted into the equation to have (33e). Again, $[I_{x,5}(1), I_{y,5}(1)]_p^T$ is formally linear about $[V_{x,5}(1), V_{y,5}(1)]^T$ with coefficient matrices denoted by $A_{p,5}$ and $B_{p,5}$.

After obtaining the linear forms for all current injections, we can combine them into the matrix representation (34). For instance, $A_1$ and $B_1$ are equal to the $A_{g,1}$ and $B_{g,1}$ respectively at generator bus 1.

$$\begin{bmatrix} I_{x,1}(1) \\ \vdots \\ I_{y,9}(1) \end{bmatrix} = \begin{bmatrix} A_1 & & \\ & \ddots & \\ & & A_9 \end{bmatrix} \begin{bmatrix} V_{x,1}(1) \\ \vdots \\ V_{y,9}(1) \end{bmatrix} + \begin{bmatrix} B_1 \\ \vdots \\ B_9 \end{bmatrix} \quad (34)$$

Finally, combining (34) with the DTs of the network equation (23), $V(1)$ is explicitly solved in (35). By recursively conducting this process, coefficients of bus voltages and state variables with any order $k$ can be obtained.

$$\begin{bmatrix} V_{x,1}(1) \\ \vdots \\ V_{y,9}(1) \end{bmatrix} = \left( \begin{bmatrix} Y_{11} & \cdots & Y_{19} \\ \vdots & \ddots & \vdots \\ Y_{91} & \cdots & Y_{99} \end{bmatrix} - \begin{bmatrix} A_1 & & \\ & \ddots & \\ & & A_9 \end{bmatrix} \right)^{-1} \begin{bmatrix} B_1 \\ \vdots \\ B_9 \end{bmatrix} \quad (35)$$

In each time step, solutions of involved variables are approximated by power series of time in (3). By performing the above process in multiple time steps, the solutions over a desired simulation range is obtained. Fig. 2a gives the transient voltage trajectories at bus 1 and bus 5 after a large disturbance using both the proposed method with $K=8$ and time step length of 0.01 s, and the TRAP-NR method with time step length of 0.001 s. Fig. 2b further provides the maximum voltage errors of all 9 buses for both methods compared with the benchmark result. It shows that the error of the proposed method is reduced by one order of magnitude compared to that of the TRAP-NR method over the entire simulation period despite the 10 times prolonged time step length.

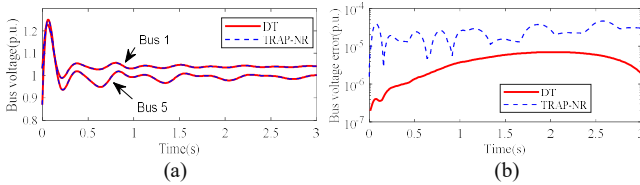

Fig. 2. Trajectories and errors of bus voltages for the 9-bus system (a) Voltage trajectories at bus 1 and bus 5 (b) Maximum voltage error of all 9 buses

### B. Accuracy and Time Performance

Both stable and unstable scenarios are simulated for the Polish system to validate the accuracy and time performance of the proposed method.

Respectively for two scenarios, Fig. 3 and Fig. 4 respectively show the transient responses of rotor angles, rotor speeds and bus voltages simulated by the proposed method. The machine 1 is selected as the reference to calculate relative rotor angles. The maximum errors of rotor angles, rotor speeds, and bus voltages compared with the benchmark results are $3.02 \times 10^{-5}$ degree, $4.27 \times 10^{-7}$ Hz, $3.33 \times 10^{-7}$ p.u. for stable scenario and $2.02 \times 10^{-5}$ degree, $3.00 \times 10^{-7}$ Hz, $2.78 \times 10^{-7}$ p.u. for unstable scenario respectively. It shows the proposed method can accurately simulate both stable and unstable contingencies in the transient stability simulation.

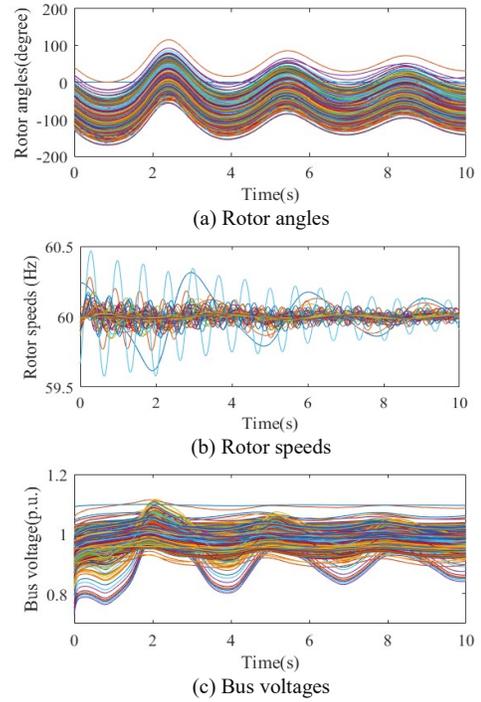

Fig. 3. Trajectories of the stable scenario for the 2383-bus system

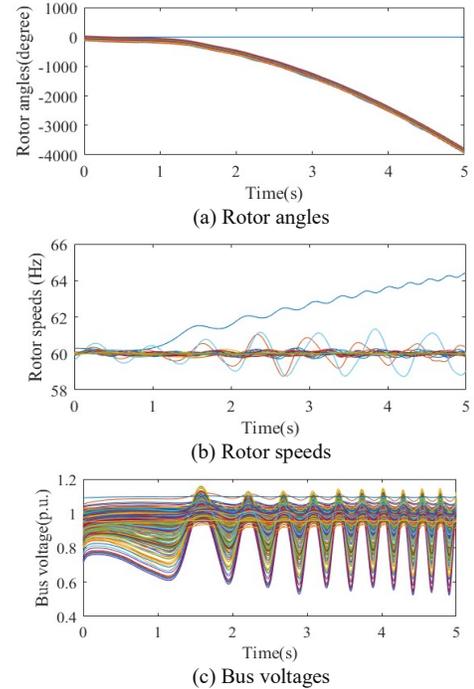

Fig. 4. Trajectories of the unstable scenario for the 2383-bus system

Table I gives the maximum errors of all state variables (including rotor angles, rotor speeds, transient and sub-transient voltages, field voltages, etc.) and bus voltages respectively, as well as the computation time per 1-second simulation. The errors of the state variables and the bus voltages using the proposed method are respectively two orders of magnitude lower and one order of magnitude lower than those using the TRAP-NR method and the ME-NR method. Also, the computation speed of the proposed method is around 10 times faster than the other two methods. These results show the proposed method is more efficient and accurate.

TABLE I
COMPARISON OF ACCURACY AND TIME PERFORMANCE

| Scenarios | Methods | Error of state variables (p.u.) | Error of bus voltages (p.u.) | Computation time (s) |
|---|---|---|---|---|
| Stable | DT | $2.69 \times 10^{-6}$ | $3.33 \times 10^{-7}$ | 18.76 |
| | TRAP-NR | $1.30 \times 10^{-4}$ | $1.10 \times 10^{-6}$ | 176.43 |
| | ME-NR | $2.63 \times 10^{-4}$ | $2.26 \times 10^{-6}$ | 191.40 |
| Unstable | DT | $1.89 \times 10^{-6}$ | $2.78 \times 10^{-7}$ | 18.85 |
| | TRAP-NR | $1.41 \times 10^{-4}$ | $1.61 \times 10^{-6}$ | 182.76 |
| | ME-NR | $2.79 \times 10^{-4}$ | $2.93 \times 10^{-6}$ | 196.02 |

Fig. 5 gives the error propagation along the simulation process for four scenarios with the time step length increased to 0.02 s, 0.05s, 0.10s and 0.20s respectively starting from the same initial states at $t$=0. It shows that the error does not accumulate much when the time step length is 0.02 s and 0.05 s. The maximum errors are in the order of magnitude of $10^{-5}$ p.u. and $10^{-3}$ p.u.. respectively. For larger time step lengths, the error becomes unneglectable when the time step length is 0.10 s and even reaches $10^5$ p.u. when the time step length is 0.20 s, indicting divergence of the solution.

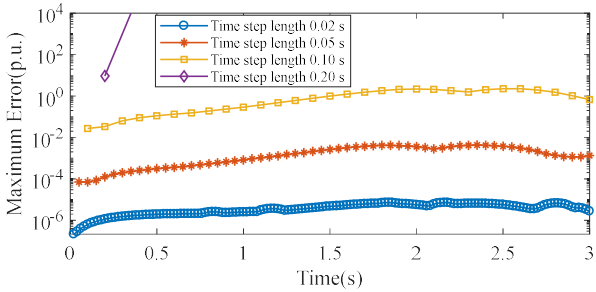

Fig. 5. Error propagation under different time step lengths

In this paper, the $K$ is determined by gradually increasing its value until the maximum error of all variables satisfies a pre-defined requirement. Table II gives the error and the computation time with different values of $K$. It shows that the errors of state variables and bus voltages are decreased when $K$ increases from 2 to 8. And, keeping increasing $K$ does not further improve the accuracy much but brings more computation burden. Therefore, $K$=8 is selected throughout the case study to meet the accuracy requirement, where the maximum error of all variables is in the order of magnitude of $10^{-6}$ p.u..

TABLE II
ACCURACY AND TIME PERFORMANCE FOR DIFFERENT VALUES OF $K$

| $K$ | Error of state variables (p.u.) | Error of bus voltages (p.u.) | Computation time (s) |
|---|---|---|---|
| 2 | $2.70 \times 10^{-2}$ | $4.91 \times 10^{-4}$ | 8.78 |
| 3 | $8.51 \times 10^{-4}$ | $2.10 \times 10^{-5}$ | 10.31 |
| 4 | $3.33 \times 10^{-5}$ | $1.82 \times 10^{-6}$ | 11.82 |
| 8 | $2.69 \times 10^{-6}$ | $3.33 \times 10^{-7}$ | 18.76 |
| 12 | $2.69 \times 10^{-6}$ | $3.33 \times 10^{-7}$ | 27.91 |

Since a large computation burden with transient stability simulation lies in solving linear equations, both sparse matrix and LU factorization techniques are implemented in this paper for the DT method, TRAP-NR method and ME-NR method. Table III compares the total number $N_{LU}$ of times of LU factorization with three methods in a 1-second simulation. It is calculated by $N_{LU}=n_{LU}\times M$, where $n_{LU}$ is the number of times of LU factorization within each time step and $M$ is the total number of time steps. Within each time step, both the TRAP-NR and the ME-NR method need to perform LU factorization for each iteration unless a so-called very dishonest NR method is applied, but the DT method only needs to perform LU factorization once. Also, the DT method only takes 1/10 of time steps of the other two methods. Therefore, the DT method can significantly reduce the number of times of LU factorization in a simulation.

TABLE III
COMPARISON OF NUMBER OF LU FACTORIZATION

| Methods | $n_{LU}$ | $M$ | $N_{LU}=n_{LU} \times M$ |
|---|---|---|---|
| DT | 1 | 100 | 100 |
| TRAP-NR | 3.004 | 1000 | 3004 |
| ME-NR | 2.060 | 1000 | 2006 |

*C. Robustness*

The robustness of the proposed method is validated in three sets of cases: 1) stable and unstable scenarios, 2) disturbances with different severities, and 3) different percentages of constant power load.

By comparing the results of stable and unstable scenarios in Table I, it is observed that the TRAP-NR method and ME-NR method are less accurate and slower in simulating the unstable scenario than in the stable scenario, but the proposed DT-based method performs almost the same in both scenarios. This is because the system states change significantly in the unstable scenario and the NR method takes more iterations to converge. At each time step, the TRAP-NR method takes averagely 3.004 iterations in the stable scenario and 3.118 iterations in the unstable scenario. For ME-NR method, it takes 2.060 and 2.132 iterations, respectively. In contrast, the proposed method does not require iterations in the solving process, thus having better robustness on unstable scenarios.

Fig. 6 gives the time performance and the average number of iterations of the NR method under different disturbances with increasing severities using the three methods. It shows the computation time of the proposed method is almost the same for different disturbances, but both the TRAP-NR method and ME-NR method take longer time when simulating larger disturbances due to the increased number of iterations.

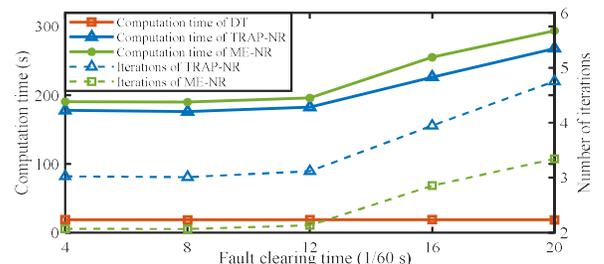

Fig. 6. Robustness against different disturbances

The time performance and the average number of iterations of the NR method under different percentages of constant power load is in Fig. 7. The higher percentage of constant power load brings stronger nonlinearity to the DAEs, thus making the NR method more difficult or fail to converge. In Fig. 7, the computation time of both the TRAP-NR method and ME-NR method increase significantly with the higher percentage of the constant power load. But the proposed method does not need iterations and its computation time is not affected much, showing it is more robust to handle the strong nonlinearity caused by constant power load.

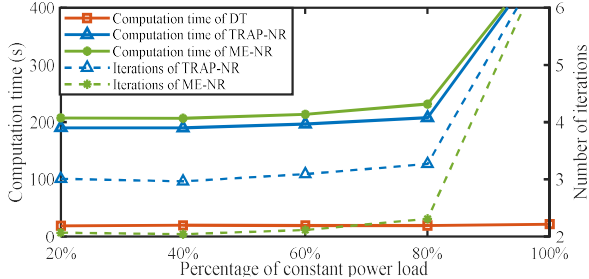

Fig. 7. Robustness against different percentages of constant power load

## VI. CONCLUSION

In this paper, a DT based non-iterative method is proposed for solving power system DAEs. Current injections and bus voltages coupled by nonlinear network equations in the original state space representation are proved to satisfy a formally linear equation in terms of their power series coefficients after DT. Benefiting from this proposition, solutions of both state variables and non-state voltages are calculated by power series of time whose coefficients are explicitly solved using the designed algorithm with no iteration. Simulation results shows the proposed method effectively reduces the computation burden compared to traditional numerical methods and demonstrates reliable time performance when solving DAEs under large disturbances or with strong nonlinearities.

## APPENDIX

**Proof of (24) in Proposition 2:** Rewrite equation (12) as follows. Define $A_g$ and $B_g$ as (36). It is easy to confirm that $A_g$ and $B_g$ do not depend on $V_x(k)$ and $V_y(k)$.

$$\begin{bmatrix} I_x(k) \\ I_y(k) \end{bmatrix} = \Gamma(k) \otimes \begin{bmatrix} E_d''(k) \\ E_q''(k) \end{bmatrix} - \sum_{m=0}^{k-1} \Lambda(k-m) \begin{bmatrix} V_x(m) \\ V_y(m) \end{bmatrix} - \Lambda(0) \begin{bmatrix} V_x(k) \\ V_y(k) \end{bmatrix}$$

$$A_g = -\Lambda(0), \quad B_g = \Gamma(k) \otimes \begin{bmatrix} E_d''(k) \\ E_q''(k) \end{bmatrix} - \sum_{m=0}^{k-1} \Lambda(k-m) \begin{bmatrix} V_x(m) \\ V_y(m) \end{bmatrix} \quad (36)$$

**Proof of (25) in Proposition 2:** To prove (25), we only need to prove each component of the ZIP load in (18)-(20) can be written into following three equations, respectively.

$$\begin{bmatrix} I_x(k) \\ I_y(k) \end{bmatrix}_z = A_z \begin{bmatrix} V_x(k) \\ V_y(k) \end{bmatrix} + B_z, \quad \begin{bmatrix} I_x(k) \\ I_y(k) \end{bmatrix}_i = A_i \begin{bmatrix} V_x(k) \\ V_y(k) \end{bmatrix} + B_i$$

$$\begin{bmatrix} I_x(k) \\ I_y(k) \end{bmatrix}_p = A_p \begin{bmatrix} V_x(k) \\ V_y(k) \end{bmatrix} + B_p$$

**Part a)** The DT of current injections of constant impedance load (18) can be easily written into above forms, by defining $A_z$, $B_z$ as (37).

$$A_z = \frac{1}{v_{t0}^2} \beta_a, \quad B_z = 0_{2 \times 2} \quad (37)$$

**Part b)** The DT of current injections of constant current load (19) is rewritten as follows.

$$\begin{bmatrix} I_x(k) \\ I_y(k) \end{bmatrix}_i = \frac{1}{v_{t0}^2} \left( \beta_b \begin{bmatrix} V_x(k) \\ V_y(k) \end{bmatrix} - \sum_{m=0}^{k-1} V_t(k-m) \begin{bmatrix} I_x(m) \\ I_y(m) \end{bmatrix}_i \right)$$

$$= \frac{1}{v_{t0}^2} \beta_b \begin{bmatrix} V_x(k) \\ V_y(k) \end{bmatrix} - \frac{1}{v_{t0}^2} \sum_{m=1}^{k-1} V_t(k-m) \begin{bmatrix} I_x(m) \\ I_y(m) \end{bmatrix}_i - \frac{1}{v_{t0}^2} V_t(k) \begin{bmatrix} I_x(0) \\ I_y(0) \end{bmatrix}_i$$

$$\triangleq A_{i,1} \begin{bmatrix} V_x(k) \\ V_y(k) \end{bmatrix} + B_{i,1} - \frac{1}{v_{t0}^2} V_t(k) \begin{bmatrix} I_x(0) \\ I_y(0) \end{bmatrix}_i$$

The third term can be further written as follows after substituting $V_t(k)$ in (17).

$$-\frac{1}{v_{t0}^2} V_t(k) \begin{bmatrix} I_x(0) \\ I_y(0) \end{bmatrix}_i = -\frac{1}{2v_{t0}^3} \left( U(k) - \sum_{m=1}^{k-1} V_t(m) V_t(k-m) \right) \begin{bmatrix} I_x(0) \\ I_y(0) \end{bmatrix}_i$$

$$\triangleq -\frac{1}{2v_{t0}^3} U(k) \begin{bmatrix} I_x(0) \\ I_y(0) \end{bmatrix}_i + B_{i,2}$$

The first term can be further written as follows after substituting $U(k)$ in (16).

$$-\frac{1}{2v_{t0}^3} U(k) \begin{bmatrix} I_x(0) \\ I_y(0) \end{bmatrix}_i = -\frac{1}{2v_{t0}^3} \{V_x(k) \otimes V_x(k) + V_y(k) \otimes V_y(k)\} \begin{bmatrix} I_x(0) \\ I_y(0) \end{bmatrix}_i$$

$$= -\frac{1}{2v_{t0}^3} \left\{ \sum_{m=1}^{k-1} V_x(m) V_x(k-m) + \sum_{m=1}^{k-1} V_y(m) V_y(k-m) \right\} \begin{bmatrix} I_x(0) \\ I_y(0) \end{bmatrix}_i$$

$$- \frac{1}{v_{t0}^3} \{V_x(0) V_x(k) + V_y(0) V_y(k)\} \begin{bmatrix} I_x(0) \\ I_y(0) \end{bmatrix}_i$$

$$\triangleq B_{i,3} + A_{i,2} \begin{bmatrix} V_x(k) \\ V_y(k) \end{bmatrix}, \text{ where } A_{i,2} = -\frac{1}{v_{t0}^3} \begin{bmatrix} I_x(0) \\ I_y(0) \end{bmatrix}_i [V_x(0), V_y(0)]$$

Finally, define $A_i$ and $B_i$ as (38). It is easy to confirm that $A_i$ and $B_i$ do not depend on $V_x(k)$ and $V_y(k)$.

$$A_i = A_{i,1} + A_{i,2}, \quad B_i = B_{i,1} + B_{i,2} + B_{i,3} \quad (38)$$

**Part c)** Equation (20) is written as follows.

$$\begin{bmatrix} I_x(k) \\ I_y(k) \end{bmatrix}_p = \frac{1}{v_{t0}^2} \left( \beta_c \begin{bmatrix} V_x(k) \\ V_y(k) \end{bmatrix} - \sum_{m=0}^{k-1} U(k-m) \begin{bmatrix} I_x(m) \\ I_y(m) \end{bmatrix}_p \right)$$

$$= \frac{1}{v_{t0}^2} \beta_c \begin{bmatrix} V_x(k) \\ V_y(k) \end{bmatrix} - \frac{1}{v_{t0}^2} \sum_{m=1}^{k-1} U(k-m) \begin{bmatrix} I_x(m) \\ I_y(m) \end{bmatrix}_p - \frac{1}{v_{t0}^2} U(k) \begin{bmatrix} I_x(0) \\ I_y(0) \end{bmatrix}_p$$

$$\triangleq A_{p,1} \begin{bmatrix} V_x(k) \\ V_y(k) \end{bmatrix} + B_{p,1} - \frac{1}{v_{t0}^2} U(k) \begin{bmatrix} I_x(0) \\ I_y(0) \end{bmatrix}_p$$

The third term can be further written as follows after substituting $U(k)$ in (16). It is easy to confirm that $A_p$ and $B_p$ defined in (39) do not depend on $V_x(k)$ and $V_y(k)$.

$$-\frac{1}{v_{t0}^2} U(k) \begin{bmatrix} I_x(0) \\ I_y(0) \end{bmatrix}_p$$

$$= -\frac{1}{v_{t0}^2} \left\{ \sum_{m=1}^{k-1} V_x(m) V_x(k-m) + \sum_{m=1}^{k-1} V_y(m) V_y(k-m) \right\} \begin{bmatrix} I_x(0) \\ I_y(0) \end{bmatrix}_p$$

$$- \frac{2}{v_{t0}^2} \{V_x(0) V_x(k) + V_y(0) V_y(k)\} \begin{bmatrix} I_x(0) \\ I_y(0) \end{bmatrix}_p$$

$$\triangleq B_{p,2} + A_{p,2} \begin{bmatrix} V_x(k) \\ V_y(k) \end{bmatrix}, \text{ where } A_{p,2} = -\frac{2}{v_{t0}^2} \begin{bmatrix} I_x(0) \\ I_y(0) \end{bmatrix}_p [V_x(0), V_y(0)]$$

$$A_p = A_{p,1} + A_{p,2}, \quad B_p = B_{p,1} + B_{p,2} \quad (39)$$

Finally, (25) is proved by defining $A_l$ and $B_l$ as (40).

$$A_l = A_z + A_i + A_p$$
$$B_l = B_z + B_i + B_p \quad (40)$$

**Proof of (30):** Rewrite (29) as follows.

$$I_{l,f}(k) = I_l(k) + dI_l(k) \otimes \Delta F(k)$$
$$= (1 + d\Delta F(0))I_l(k) + dI_l(0)\Delta F(k) + d\sum_{m=1}^{k-1} I_l(m)\Delta F(k-m)$$

Since $I_l(k) = A_l V_l(k) + B_l$ has been proved for the ZIP load model, the above equation is further rewritten as follows.

$$I_{l,f}(k) = (1 + d\Delta F(0))(A_l V_l(k) + B_l) + dI_l(0)\Delta F(k)$$
$$+ d\sum_{m=1}^{k-1} I_l(m)\Delta F(k-m)$$
$$= (1 + d\Delta F(0))A_l V_l(k) + dI_l(0)\Delta F(k) + (1 + d\Delta F(0))B_l + d\sum_{m=1}^{k-1} I_l(m)\Delta F(k-m)$$

Finally, by defining $A_l', A_f, B_l'$ in (41), the linear relationship (30) is satisfied, where $A_l', A_f, B_l'$ do not depend on $V_x(k)$ and $V_y(k)$.

$$A_l' = (1 + d\Delta F(0))A_l; A_f = dI_l(0);$$
$$B_l' = (1 + d\Delta F(0))B_l + d\sum_{m=1}^{k-1} I_l(m)\Delta F(k-m) \quad (41)$$

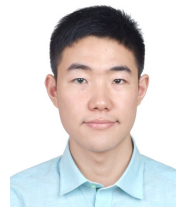

**Yang Liu** (S'17) received the B.S. degree in energy and power engineering from Xi'an Jiaotong University, China, in 2013, and the M.S. degree in power engineering from Tsinghua University, China, in 2016, respectively. He is currently pursuing the Ph.D. degree at the Department of Electrical Engineering and Computer Science, University of Tennessee, Knoxville, USA. His research interests include power system simulation, stability and control.

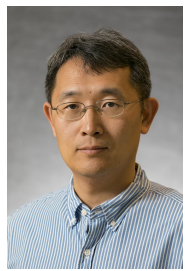

**Kai Sun** (M'06–SM'13) received the B.S. degree in automation in 1999 and the Ph.D. degree in control science and engineering in 2004 both from Tsinghua University, Beijing, China. He is an associate professor at the Department of EECS, University of Tennessee, Knoxville, USA. He was a project manager in grid operations and planning at the EPRI, Palo Alto, CA from 2007 to 2012. Dr. Sun serves in the editorial boards of IEEE Transactions on Power Systems, IEEE Transactions on Smart Grid, IEEE Access and IET Generation, Transmission and Distribution.